\documentstyle[12pt,epsfig]{article}                        
\newcommand{\ber}{\begin{eqnarray}}
\newcommand{\eer}{\end{eqnarray}}
\newcommand{\bea}{\begin{equation}}
\newcommand{\eea}{\end{equation}}
\newcommand{\del}{\partial}
\begin{document}
\title{\bf Finite Temperature Excitations of a trapped Bose gas by Feynman-Kac
path integral approach} 
\author {\bf S. Datta \\
Department of Theoretical Physics\\
2A and 2B Raja S.C.Mullick Road\\
Indian Association for the Cutivation of Science\\
Jadavpur, Kolkata 700 032, India \\}
\maketitle
\begin{abstract}

We present results from a detailed Quantum Monte Carlo study of BEC applied 
to JILA experiment[Jin et al, Phys. Rev. Lett.$\bf 78$,764,1997][1]. This is the 
first Monte Carlo approach ( based on Feynman-Kac path integral method) to the 
above problem where good  qualitative agreement is found 
for both the lowest lying  $ m=2 $ and $ m=0 $ mode. We found an
upward shift of the experimental data for $ m=0 $ mode at around 
$ T=0.7 T_0 $ ($T_0$ is defined as the predicted BEC transition temperature 
for a harmonically confined ideal gas)  when the effect of noncondensate was considered. 
\end{abstract}
\newpage
\section{Introduction}
After the experimental realization of Bose Einstein Condensation in alkali 
vapors in 1995[2], and subsequent experiments pertinent to temperature 
dependence of frequencies and damping rate[1], there have been a lot of 
theoretical studies [3-9] to explain the experimental observations in 
connection with temperature dependent frequency shifts corresponding to 
different angular momenta, m=0 and m=2 modes in particular[2]. Results have 
been reported in which theoretical data agreed well with experimental values 
for m=0 mode showing an upward trend of frequencies with rise in temperature
[4,9]. But in all cases the agreement is rather poor when it comes to m=2 
mode. There is an agreement up to  $ T=0.6 T_0 $, beyond which frequencies rise 
with increase in temperature deviating from the downward trend of experimental 
data. In this article, we would like to report a diffusion Monte Carlo study 
of the frequency shifts of m=2 and m=0 modes in a dilute gas of $Rb^{87}$.  
In our non mean field study, we see agreement with experimental study (Fig 3a of Ref 2]for $m=2$ 
mode all the way to $ T=0.9T_0 $ (Fig 6).
 When we consider the dynamics of the thermal cloud separately, the upward 
shift ( Fig 7) at $ T=0.7 T_0 $ which is similar to JILA experiment is  
observed for $m=0$ mode. This agrees with the results  obtained from the 
revised gapless theory of Morgan[6,7].

The dynamical behavior of dilute alkali BECs at T=0 can be well described by 
Gross-Pitaevskii eqn(GPE)[10].
\\
\bea
i\hbar \frac {\del \phi(\vec{r},t)}{\del t}
=[-\frac{{\hbar}^2}{2m_{Rb}} \frac {\Delta}{2}+V_{ext}(\vec{r})
+g|{\phi(\vec{r},t)}|^2]\phi(\vec{r},t)
\eea
where $ g=\frac{4\pi{\hbar}^2a}{m_{Rb}}$, 'a' is the scattering length and
'$m_{Rb}$' is the mass of Rb atom.
But it seems to be inadequete at finite temperatures.
The total density of the atoms is related to BEC density and
normal component as follows: At T=0 the normal component is not equal to zero
in the interacting case and is referred to as 'quantum depletion'. At finite
temperature, thermal atoms also contribute to the normal component. Since mean
field wavefunctions do not  account for the normal component, it gives accurate
energy spectrum if depletion is small[11]. Near $T_0$, the quantum depletion
becomes significant and mean field treatment breaks down.  

The other mean field theories which have been used so far, are based on HF and 
HFB-Popov equations [12] and  break down near $ T_0 $ as its effective single 
particle spectrum always displays a gap. In 1998, a self consistent 
gapless non-divergent theory[5] was developed and a closed 
set of coupled equations were solved numerically. In this analysis, 
only the dynamics of condensate was considered and downshift of data was
observed for both $m=0$ and $m=2$ mode. Subsequently, with the more
sophisticated theory of Morgan[6] an upward shift at $T=0.6 T_0$  was 
achieved[7,8]
Analytic expressions[13] for temperature dependent frequencies were obtained 
for m=0 and m=2 by time dependent variational technique. 

Eventhough Ref[7] has the best agreement with JILA data till date, solving coupled partial differential equations numerically is not an easy task. 
The chief purpose of this paper is to go beyond mean field theory with a 
comparatively simpler numerical procedure which would work at all 
temperatures. We propose to explore finite temperature aspect of BEC by 
quantum nonperturbative technique, namely Feynman- Kac (FK)[14-16] procedure. 
To be precise, we use Generalised Feynman-Kac  ( GFK ) method[17] to make the 
rate of convergence faster. Since Quantum Monte Carlo methods are 
computationally expensive, we are simulating only 2000 interacting atoms at 
this moment. Increasing number of interacting atoms would change our results 
quantitatively,not qualitatively. 
This paper is organized as follows : 
In Sec 2, we discuss the path integral technique at zero and finite temperature
as a many body technique, the Schroedinger formulation of Rb condensate and 
noncondensate, fundamental concepts of BEC and finite temperature 
excitations. In Sec 3, we discuss the numerical procedure. In Sec 4, we 
present all the numerical results pertinent to energies and frequencies at 
different temperature. 
Finally in Sec 5, we summarize our results. 

\section{ Theory }
To connect Feynman-Kac or Generalized Feynman Kac ( GFK ) to other many body 
techniques our numerical procedure ( GFK )[18-19] has a straightforward 
implementation to Schroedinger's wave mechanics. Since at low temperature the 
de Broglie wavelength of the atoms become appreciable, we do a full quantum 
treatment. GFK is essentially a path integral technique with trial functions 
for which operations of the group of the wave function keep points in the 
chosen nodal region, provide an upper bound for the lowest state energy of that
symmetry. The nodal region with the lowest energy serves as a least upper 
bound. If the nodal region has exact nodal structures of the true wave function the random walk is exact in the limit scale, time for walk, and number of walks get arbitrarily large. To calculate energy we approximate an exact solution 
( i.e.,the GFK representation of it ) to the Schroedinger's equation, whereas 
most of the other numerical procedures approximate a solution to an approximate 
Schroedinger equation. From the equivalence of the imaginary time
propagator and temperature dependent density matrix, finite temperature
results can be obtained from the same zero temperature code by running it
for finite time.  So from all these aspects, Generalised Feynman-Kac method
turns out to be a potentially good candidate as a sampling procedure for Bose
gases at all temperatures. Next we consider the Feynman-Kac 
formalism and then show how it can be modified to get the Generalized 
Feynman-Kac version of it.

\newpage
\subsection {  Path integral Theory at T=0}
\subsubsection { Feynman-Kac Path integretion }
For the Hamiltonian $H=-\Delta/2+V(x)$ consider the initial value problem
\ber
i\frac{\del u(t,x)}{\del t}& =& (-\frac{\Delta}{2}+V)u(t,x)\nonumber\\
& & u(0,x)=f(x)
\eer
with $x \in  R^d$ and $u(0,x)=1$. The solution of the above equation can
be written in Feynman-Kac representation as
\bea
u(t,x)=E_xexp[-\int_0^t V(X(s))ds]
\eea
where X(t) is a Brownian motion trajectory and $E_x$ is the average value of the
exponential term with respect to these trajectories. The lowest energy eigenvalue for a  given symmetry can be obtained from the large deviation
principle of Donsker and Varadhan [20],
\bea
\mu=-\lim_{t\rightarrow \infty} \frac{1}{t}ln[E_xexp[-\int_0^t V(X(s))ds]]
\eea
The above formalism is valid for any arbitrary dimension d  
( for a system of N particles in three dimensions $d=3N$). 
Generalizations of the class of potential functions for which Eqns. 3 and 4
are valid are given by Simon[21] and include most physically interesting
potentials, positive or  negative, including, in particular, potentials
with $1/x$ singularities. It can be argued that the functions determined
by Eq(3) will be the one with lowest energy of all possible functions
independent of symmetry. Restrictions on allowed Brownian motions
must be imposed to get a solution of the desired symmetry if it is not the
lowest energy solution for a given Hamiltonian. Since the above energy formula 
gives the lowest energy corresponding to any symmetry, the same formula can be 
used  to calculate ground and excited states of a quantum mechanical system.
Although other interpretations are interesting, the simplest is that the Brownian motion distribution is just a useful mathematical construction which allows 
one to extract the  physically relevant quantities, the ground and excited 
state energy of a quantum mechanical system. In numerical implementation of 
Eq(4) the 3N dimensional Brownian motion is replaced by 3N independent, 
properly scaled one dimensional random walks as follows. For a given time t 
and integers n and l define [18] the vector in $R^{3N}$
\ber
W(l)\equiv W(t,n,l)
& = & ({w_1}^1(t,n,l),{w_2}^1(t,n,l),{w_3}^1(t,n,l)....\\ \nonumber
&   &                  .......{w_1}^N(t,n,l){w_2}^N(t,n,l){w_3}^N(t,n,l)
\eer
where
\bea
{w_j}^i(t,n,l)=\sum^l_{k=1}\frac{{\epsilon}^i_{jk}}{\sqrt n}
\eea
with ${w_j}^i(0,n,l)=0$
for $i=1,2,....,N$;$j=1,2,3$ and $l=1,2,.....,nt$. Here $\epsilon $ is
chosen independently and randomly with probability P for all i,j,k such that
$P({\epsilon}^i_{jk}=1)$=$P({\epsilon}^i_{jk}=-1)$=$\frac{1}{2}$. It is known 
by an invariance principle[22] that for every $\nu$ and W(l)
defined in Eq(5)
                                                                                
\ber
\lim_{n\to\infty}P(\frac{1}{n}\sum^{nt}_{l=1}V(W(l)))\leq \nu \\ \nonumber
 =  P( \int\limits^t_0 V( X(s))ds\leq\nu
\eer
Consequently for large n,
\ber
P[ \exp(- \int\limits^t_0 V(X(s))ds)\leq\nu ] \\ \nonumber
 \approx  P [\exp(-\frac{1}{n}\sum^{nt}_{l=1}V(W(l)))\leq \nu]
\eer
By generating $N_{rep}$ independent replications $Z_1$,$Z_2$,....$Z_{N_{rep}}$ of
\bea
Z_m=\exp(-(-\frac{1}{n}\sum^{nt}_{l=1}V(W(l)))
\eea
and using the law of large numbers, $(Z_1+Z_2+...Z_{N_{rep}})/N_{rep}=Z(t)$
is an approximation to Eq(3)
\bea
\mu\approx -\frac{1}{t}logZ(t)
\eea
Here $W^m(l), m=1,2,N_{rep}$ denotes the $m^{th}$ realization of W(l) out of
$N_{rep}$ independently run simulations. In the limit of large t and $N_{rep}$
this approximation approaches an equality, and forms the basis of a
computational scheme for the lowest energy of a many particle system with
a prescribed symmetry.                                                          In dimensions higher than 2, the trajectory x(t) escapes to infinity with
probability 1 . As a result, the important regions of the potential are
sampled less and less frequently and the above equation converges slowly.
Now to speed up the convergence we use Generalized Feynman-Kac (GFK) method.
\subsubsection { Generalized Feynman Kac path integretion }
To formulate the generalized Feynman-Kac method we first rewrite the
Hamiltonian as $H=H_0+V_p$, where
$ H_0=-{\Delta}/2+{\mu}_T+{{\Delta}{\psi}_T}/{2{\psi}_T} $ and
$V_p=V-({\mu}_T+\Delta{\psi}_T/2{\psi}_T)$.
Here ${\psi}_T$ is a twice differentiable nonnegative reference function
and $H{\psi}_T={\mu}_T{\psi}_T$. The expression for the energy can now be
written as
\bea
\mu={\mu}_T-\lim_{t\rightarrow \infty}
 \frac{1}{t}ln[ E_xexp[-\int_0^t V_p(Y(s))ds]]
\eea
where Y(t) is the diffusion process which solves the stochastic differential
equation
\bea
dY(t)=\frac{{\Delta}{\psi}_T(Y(t))}{{\psi}_T(Y(t))}dt+dX(t)
\eea
The presence of both drift and diffusion terms in this expression
enables the trajectory Y(t) to be highly localized. As a result, the important
regions of the potential are frequently sampled and Eq (11) converges rapidly.\\

\subsection { Path integral theory at finite temperature}
                                                                           
The temperature dependence comes from the realization that
the imaginary time propagator $k(2,1)$ is identical to the temperature
dependent density matrix $\rho(2,1)$ if $t\Rightarrow\beta=  1/T$
holds.
        
This becomes obvious when we consider the eqs[23]
\bea
-\frac{\del k(2,1)}{\del t_2}= H_2k(2,1)
\eea
and
\bea
-\frac{\del \rho}{\del \beta}=H_2 \rho(2,1)
\eea
and compare
\bea
k(2,1)=\sum_i\phi_i(x_2){\phi_i}^*(x_1)e^{-(t_2-t_1)E_i}
\eea
and
\bea
\rho(2,1)=\sum_i\phi_i(x_2){\phi_i}^*(x_1)e^{-\beta E_i}
\eea
For Zero temperature FK we had to extrapolate to $t\Rightarrow \infty$.
For finite run time t in the simulation, we have finite temperature results.
In this section we show how we change our formalism to go from zero to finite
temperature. We begin with the definition of finite temperature.
A particular temperature 'T' is said to be finite if
$\Delta E < kT$ holds.
The temperature dependent density matrix can be written in the following form
\bea
\rho(x,x^{\prime},\beta)={\rho}^{(0)}(x,x^{\prime},\beta)\nonumber\\
\times < exp[-\int_{0}^{\beta}V_p[X(s)]ds]>_{DRW}
\eea
The partition function can be recovered from the above as follows:
\bea
\int{\rho(x,x,\beta)} dx=\int{{\rho}^{(0)}(x,x,\beta)}dx
\times  < exp[-\int_{0}^{\beta}V_p[X(s)]ds]>_{DRW}
\eea
In the usual notation, the above equation reads as
\bea
Z(x,\beta)=Z^{0}(x,\beta)\times  < exp[-\int_{0}^{\beta}V_p[X(s)]ds]>_{DRW}
\eea
At finite temperature thus free energy can be written as
\bea
F=
-ln Z(x,\beta)/\beta=-{ln Z^{0}(x,\beta)}/\beta-{ln  < exp[-\int_{0}^{\beta}
V_p[X(s)]ds]>_{DRW}}/\beta
\eea
\newpage
\subsection{ Schroedinger Formalism for Rb condensate at T=0 }
In the JILA experiment different frequency modes are labeled by their angular 
momentum projection on the trap axis. In cylindrical symmetry, $m=2$ mode is an 
uncoupled one and there are two coupled oscillations for $m=0$ mode [24]. 
As a matter of fact Stringari [25] showed that $m=0$ mode is a coupled 
oscillation of a quadrupolar surface oscillation and a monopole. 
In the noninteracting case, these two modes are degenerate with
$\omega/{\omega}_x=2$ .\\
We choose to work in the cylindrical coordinates as  
the original experiment had an axial symmetry, cylindrical coordinates are the 
natural choices for this problem. We consider a cloud of N atoms interacting 
through repulsive potential placed in a three dimensional harmonic oscillator 
potential. At low energy the stationary state for the condensate can be 
represented as 
\bea
[-{\Delta}/2+V_{int}+V_{trap}]\psi_0(\vec{r})
={\mu}_{c}\psi_0(\vec{r})
\eea

\bea
[-{\Delta}/2+V_{int}+\frac{1}{2} \sum_{i=1}^N [{x_i}^2
 +  {y_i}^2+{\lambda_a}{z_i}^2]]\psi_0(\vec{r}) 
={\mu}_{c}\psi_0(\vec{r}) 
\eea
where $\frac{1}{2} \sum_{i=1}^N [{x_i}^2
 +  {y_i}^2+{\lambda_a}{z_i}^2]$ is the anisotropic 
potential with anisotropy factor ${{\lambda}_a}=\frac{{\omega}_z}{{\omega}_x}$.
Now 
 \bea
 V_{int}=V_{Morse}=
\sum_{i,j} V(r_{ij})=\sum_{i<j}D[e^{-\alpha(r-r_0)}(e^{-\alpha(r-r_0)}-2)]
 \eea

In the above potential '$r_0$' is the location of the well minimum and 
'$\alpha$'
is the width of the Morse potential.
The above Hamiltonian is not separable in spherical polar coordinates beacause 
of the anisotropy. In cylindrical coordinates the noninteracting part behaves 
as a system of noninteracting harmonic oscillators and can be writtem as 
follows :
\ber
& & [-\frac{1}{2 \rho}\frac{\partial}{\partial \rho}(\rho\frac{\partial}
{\partial \rho}) -\frac{1}{{\rho}^2}\frac{{\partial}^2}{\partial {\phi}^2}
-\frac{1}{2}\frac{{\partial}^2}{\partial{z}^2}\nonumber \\
& & +\frac{1}{2}({\rho}^2+{\lambda_a}^2z^2)]\psi_0(\rho,z) \nonumber \\
& = & {\mu}_{c} {\psi}_0(\rho,z)
\eer
The  energy '$\mu$' of the above equation can be calculated exactly which is
\bea
 {\mu}_{n_{\rho} {n_z} m}
=(2n_{\rho}+|m|+1)+(n_z+1/2)\lambda
\eea
In our guided random walk we use the noninteracting solution of
Schroedinger equation as the trial function as follows [26]:
\bea
{\psi}_{n_{\rho}{n_z}m}(\vec{r})\simeq \exp^{\frac{-z^2}{2}}H_{n_z}(z)\times
e^{im\phi}{\rho}^me^{-{{\rho}^2}/2}{L_{n_{\rho}}}^{m}({{\rho}^2})
\eea
\subsection {\bf The effect of noncondensate}
In the case of noncondensate the system can be considered as a thermal gas.
To calculate noncondensate energy and density we need to study the effect of
noncondensate explicitly and consider
the following stationary state for the thermal gas.
\bea
[-{\Delta}/2+2V_{int}+V_{trap}]\psi_j(\vec{r})
={\mu}_{nc}\psi_j(\vec{r})
\eea
\bea
[-{\Delta}/2+2V_{int}+\frac{1}{2} \sum_{i=1}^N [{x_i}^2
 +  {y_i}^2+\lambda{z_i}^2]\psi_{j}](\vec{r})
={\mu}_{nc}\psi_{j}(\vec{r})
\eea
The basis wavefunction ${\psi}_{j}$ which describes the noncondensate should
be chosen in such a way that it is orthogonal to ${\psi_0}$ as in Eq.(11)
The most common way to achieve an orthogonal basis in Schroedinger prescription
is to consider the dynamics of noncondensate in an effective potential[6,27]
$V_{eff}=V_{trap}+2V_{int}$. The factor 2 represents the exchange
term between two atoms in two different states. The energy in the case of
lowest lying modes  then corresponds to $\mu={\mu}_c+{\mu}_{nc}$. One can calculate
the  ${\mu}_{nc}$ using the same parameters as discussed in Sec 3.1. 
\subsection{  Fundamentals of BEC }                                            
Even though the phase of Rb vapors at T=0 is certainly solid, Bose condensates 
are preferred in the gaseous form over the liquids and solids because 
at those higher densities interactions are complicated and hard to deal with 
on an elementary level. They are kept metastable by maintaining a very low 
density. For alkali metals, $\eta$, the ratio of zero point energy and 
molecular binding energy lies between $10^{-5}$ and $10^{-3}$. According
to the theory of corresponding states[28] since for the T=0 state of alkali 
metals, $\eta$ exceeds a critical value 0.46, the  molecular binding energy 
dominates over the zero point motion and they condense to solid phase. 
 But again the life time of a gas is limited by three body recombination rate 
which is proportional to the square of the atomic density. 
It gets suppressed at low density. Magnetically trapped alkali vapors can be 
metastable depending on their densities and lifetimes. 
So by keeping the density low only two body collisions are allowed as a result 
which dilute gas approximation [29] still holds for condensates which 
tantamounts to saying $na^3<<1$ (a is the scattering length of s wave).
Now defining $n=N/V=r_{av}^{-3}$ as a mean distance between the atoms 
( definition valid for any temperature ), the dilute gas condition reads as
$a<<r_{av}$ and zero point energy dominates (dilute limit). In the dense limit,
for $a \approx r_{av} $ on the other hand the interatomic potential dominates.
The gas phase is accomplished by reducing the material density through 
evaporative cooling.  \\
\subsection{Finite temperature Excitations : }
Finite temperature excitation spectrum is obtained by using the path integral 
formalism used in Section 2.2. In our analysis, we first assume that the 
condensate oscillates in a static thermal cloud. There are no interactions 
between the condensate and the thermal cloud. The principal effect of finite 
temperature on the excitations is the depletion of condensate atoms. We want to
calculate the collective excitations of Bose Einstein condensates corresponding
to JILA Top experiment ( m=2 and m=0 mode). Eventually for $ m=0 $ mode ,
we consider the effect of thermal cloud separately.\\
{\bf Condensation fraction and Critical temperature :}
In the noninteracting case for a harmonic type external force the theoretical
prediction for condensation fraction is
\bea
N_{0}/N=1-(T/T_0)^3
\eea
Critical temperature can be defined as
\bea
T_c=\frac{0.94 \times \hbar \bar{\omega}N^{1/3}}{k_B}
\eea
\bea
\bar{\omega}=({\omega_{\rho}}^2\omega_z)^{1/3}
\eea
From Eq.(29), we see that as temperature increases, condensation fracton 
decreases in the noninteracting case. Interaction lowers the condensation 
fraction for repulsive potentials. Some particles always leave the trap 
because of the repulsive nature of the potential and moreover, 
if temperature is increased further, more particles will fall out of the trap 
and get thermally distributed. This decrease in condensation fraction 
eventually would cause the shifts in the critical temperature. 
We would observe this in Section 4.2 (Fig. 4 ). Earlier this was done by
W. Krauth[30] for a large number of atoms by path integral Monte Carlo method.
In our analysis, we denote '$T_0$' as transition temperature following Ref[2]
\newpage
\section{ Numerical procedure }
\subsection{ Dilute limit }
In the dilute limit and at very low energy only binary collisions are possible
and no three body recombination is allowed. In such two body scattering at
low energy first order Born approximation is applicable and the interaction
strength 'D' can be related to the single tunable parameter of this problem,
the s-wave scattering length 'a' through the relation given below. This single
parameter can specify the interaction completely without the details of the
potential in the case of pseudopotentials. We use Morse potential because it
has a more realistic feature of having a repulsive core at $r_{ij}=0$ than other
model potentials. Secondly, using this realistic potential allows us to 
calculate the energy spectrum exactly as opposed to the case of $\delta$ 
function potential where it is calculated perturbatively[31]. In our case the 
interaction strength depends on two more additional parameters, $ r_0 $ and 
$ \alpha $.
\bea
a=\frac {mD}{4 \pi {\hbar}^2}\int V(r)d^3r
\eea
It is worth mentioning over here that instead of actual scattering length we 
use the Born approximation to it. Since we are dealing with a case of low
energy and low temperature it is quite legitimate to use the above expression 
as a trickery to calculate the strength of Morse interaction[32]. 
 As a matter
of fact in Ref[33] the author has justified using Eq.(32) for a $\delta $
function potential. So if it is justified to do it for $\delta$ function 
potental it is even more justified to do so for Morse potential which
is finite and short-ranged.   
  
The Morse potential for dimer of rubidium can be defined as
 \bea
\sum_{i,j} V(r_{ij})=\sum_{i<j}D[e^{-\alpha(r-r_0)}(e^{-\alpha(r-r_0)}-2)]
 \eea
where $\alpha$ is the depth of the Morse potential.
Using the above potential
\bea
 D=\frac{4{\hbar}^2a{\alpha}^3}{m e^{\alpha r_0}(e^{\alpha r_0}-16)}
\eea
The Hamiltonian for Rb gas with an asymmetric trapping potential and
Morse type mutual interaction can be written as
\ber
& &[ -{\hbar}^2/2m
\sum_{i=1}^N {\nabla^{\prime}_i}^2
+ \sum_{i,j}V(r^{\prime}_{ij}) \nonumber \\
& &  +\frac{m}{2}({\omega_x}^2\sum_{i=1}^N {x^{\prime}_i}^2
+{\omega_y}^2\sum_{i=1}^N {y^{\prime}_i}^2+{\omega_z}^2\sum_{i=1}^N
{z^{\prime}_i}^2)]\psi(\vec{r}^{\prime}) \nonumber \\
& = &\mu\psi(\vec{r}^{\prime})
\eer
The above Hamiltonian can be rescaled by substituting
$\vec{r}^{\prime}=s\vec{r}$ and $\mu={\mu}_0U$ as
\ber
& &[ -\frac{{\hbar}^2}{2ms^2}
\sum_{i=1}^N {\nabla_i}^2
 +  \sum_{i<j}\frac{4{\hbar}^2a{\alpha}^3}{m s^3
e^{\alpha r_0}(e^{\alpha r_0}-16)}[e^{-\alpha(\vec{r_{ij}}-r_0)}(e^{-\alpha
(\vec{r_{ij}}-r_0)}-2)] \nonumber \\
& & + \frac{ms^2}{2}({\omega_x}^2\sum_{i=1}^N {x_i}^2
 + {\omega_y}^2\sum_{i=1}^N {y_i}^2+
{\omega_z}^2\sum_{i=1}^N
{z_i}^2)]\psi(\vec{r}) \nonumber \\
& = & {\mu}_0 U\psi(\vec{r})
\eer
\ber
& & [ \frac{1}{2}\sum_{i=1}^N {\nabla_i}^2
-4\frac{a{\alpha}^3}{se^{\alpha r_0}(e^{\alpha r_0}-16)}\sum_{i<j}
[e^{-\alpha(r_{ij}-r_0)}(e^{-\alpha(r_{ij}-r_0)}-2)] \nonumber \\
& & -\frac{m^2{\omega_x}^2s^4}{2\hbar^2}
\sum_{i=1}^N ({x_i}^2
 + \frac{{\omega_y}^2}{{\omega_x}^2} {y_i}^2+
\frac{{\omega_z}^2}{{\omega_x}^2}{z_i}^2)]\psi(\vec{r})\nonumber \\
 & = & -{\mu}_0 \frac{Ums^2}{\hbar^2}\psi(\vec{r})
\eer
Let $\frac{m^2{\omega_x}^2s^4}{\hbar^2}=1\Rightarrow
s^2= \frac{\hbar}{m\omega_x}$ is the natural unit of length. Let
$\frac{Ums^2}{\hbar^2}=1\Rightarrow U=\frac{\hbar^2}{ms^2}=\hbar\omega_x$
is the natural unit of energy.
Then the standard form of the equation  becomes
\ber
& & [\frac{1}{2}\sum_{i=1}^N {\nabla_i}^2
-\sum_{i<j}4\frac{a{\alpha}^3}{se^{\alpha r_0}(e^{\alpha r_0}-16)}\sum_{i<j}
[e^{-\alpha(r_{ij}-r_0)}(e^{-\alpha(r_{ij}-r_0)}-2)] \nonumber \\
& & -\frac{1}{2} \sum_{i=1}^N ({x_i}^2
 + \frac{{\omega_y}^2}{{\omega_x}^2} {y_i}^2+
\frac{{\omega_z}^2}{{\omega_x}^2}{z_i}^2)]\psi(\vec{r}) \nonumber\\
& = & - {\mu}_0 \psi(\vec{r})
\eer
With   $\omega_x=\omega_y=\frac{\omega_z}{\sqrt \lambda}$, the above eqn
becomes,
\ber
& & [\frac{1}{2}\sum_{i=1}^N {\nabla_i}^2
-4\frac{a{\alpha}^3}{se^{\alpha r_0}(e^{\alpha r_0}-16)}\sum_{i<j}
[e^{-\alpha(r_{ij}-r_0)}(e^{-\alpha(r_{ij}-r_0)}-2)] \nonumber \\
& & -\frac{1}{2} \sum_{i=1}^N [{x_i}^2
 +  {y_i}^2+\lambda{z_i}^2]\psi(\vec{r}) \nonumber\\
& = & - {\mu}_0 \psi(\vec{r})
\eer
                                                                                
\ber
& & [\frac{1}{2}\sum_{i=1}^N {\nabla_i}^2
-\gamma\sum_{i<j}
[e^{-\alpha(r_{ij}-r_0)}(e^{-\alpha(r_{ij}-r_0)}-2)] \nonumber \\
& & -\frac{1}{2} \sum_{i=1}^N [{x_i}^2
 +  {y_i}^2+\lambda{z_i}^2]\psi(\vec{r}) \nonumber\\
& = & - {\mu}_0 \psi(\vec{r})
\eer
Now for $\alpha=.29$ and $r_0=9.758$ (both in oscillator units)[32].
We have checked that for these choice of parameters, Morse solution is
extremely good. $ a=52 \times 10^{-10}$ cm, $s=12\times 10^{-7} cm$, 
the interaction strength $\gamma$ is given by
\bea
\gamma=4\frac{a{\alpha}^3}{se^{\alpha r_0}(e^{\alpha r_0}-16)}
=2.64 \times 10^{-5}
\eea
For mean field calculation the value of interaction strength was taken to be
$ 4.33 \times 10^{-3}$. For this problem we are interested 
in the limit $\gamma << 1 $. The case $\gamma >> 1 $ is usually known as the
Thomas Fermi limit. For  $ \gamma= 2.64 \times 10^{-5}$, the
eigenvalue equation reduces to a minimally perturbed system of d dimensional
anisotropic oscillator where $d=3N $ and N is the number of particles.
The whole concept of bound states of Morse dimers is outside the range of
this limit, so the nonexistence of two-body bound states is ensured by 
choosing the above parameters. 

Even though  $\gamma << 1 $,  we solve the
eigenvalue eqn nonperturbatively with Generalized Feynman-Kac  procedure.
Energies and frequencies at zero temperature are obtained by solving Eq.
(4) and using Eq.(11). To calculate the analogous quantities at finite 
temperature we use Eq.(20). We can get the energy of both condensate and 
noncondensate using Eq.(4)
 Eq.(20 ) by generating a  large number of paths and then averaging 
the results for all the paths. Since original Feynman-Kac method [14,15] is 
computationally inefficient we incorporate  importance sampling in our 
random walk and use trial function of the form given in Eq.(26)        

{\bf Evaluation of temperature dependent mode frequencies :}
Following the prescription in [4], we see that for a fixed 'N' 
relationship $\mu(N_0,T)\simeq\mu(N,T=0)(\frac{N_0}{N})^{2/5}$ or equivalently
${N_0}/{N}=[\frac{\mu(N_0,T)}{\mu(N,T=0)}]^{5/2}$
 generates the condensation fraction ${N_0}/N$ as a function of
time. One can generate this from experiments also. From the thermodynamical
limit we get $N_0$ as a function of time and run our zero temperature code
with same number of  $N_0$ as the dynamics of the finite temperature
condensate are essentially the same as those of a zero temperature condensate
with the same value of ${N_0}$. This is called method I. The other way to 
calculate energy is to run the code for different simulation times 
corresponding to different temperatures as time is defined as 
inverse temperature. This is 
identified as method II. $\mu(N_0,T) $ and $\mu(N_0,T=0)$ are calculated using 
$\mu=\frac{1}{N}({\mu}_{kin}+{\mu}_{ho}+2{\mu}_{int})$ and ${\mu}_{kin}$,
${\mu}_{ho}$ and ${\mu}_{int}$ are calculated as described in Ref[25].
Later in Fig.5 (our data)  of Section 4.2, 
we see the effects of interaction on the condensation fraction. 

\newpage
\section{Results}
\subsection{ Excitation spectra at T=0 for different symmetries}
We chose $Rb^{87}$ as an example of a weakly interacting dilute Bose
gas as in Ref[29]. We simulate 2000 Rb atoms interacting via the Morse 
potential.
We choose $a=52 \times 10^{-10}$cm, length scale s of the problem as 
$12 \times 10^{-7} $cm. In the following table, we show the ground state 
energy of the particle. With repulsive interactions, the Energy/particle 
increases with the incresae in number of particles in the trap [Fig. 1-2] 
whereas the energy gap between the different symmetry states decreases as 
evident from Fig. 3. However we see a different trend [Fig. 4] 
for $m=0$ mode where
the excitation frequencies increase with the increase in number of atoms.
This agrees with the Hartree-Fock spectrum in the Fig 2 of Ref[33].  
 
\begin{table}[h!]
\begin{center}
\caption {\bf Results for the ground state energy of 2000  $Rb^{87}$ atoms 
in a trap with ${\omega}_x={\omega}_y=1,\lambda={\omega}_z=\sqrt 8$ in the  
interacting case ; The table shows how energy varies with the number of 
particles in the Gross Pitaevski(GP) case [34] and GFK method. }. 
\vskip 0.5cm
\begin{tabular}{|c|c|c|c} 
\hline
N & E/N(GP) & E/N(GFK) \\ 
\hline
1 & 2.414 & 2.414 213\\
100 &2.66      &   2.52230(5)\\
200 & 2.86     &  2.63141(1) \\
500 & 3.30     &  2.9588(4) \\
1000 &3.84      &  3.5047(4) \\
2000 & 4.61     & 4.5962(3)\\
\hline
\end{tabular}
\end{center}
\end{table}
\newpage
\begin{figure}[h!]
\centering
\epsfxsize=3.8in{\epsfbox{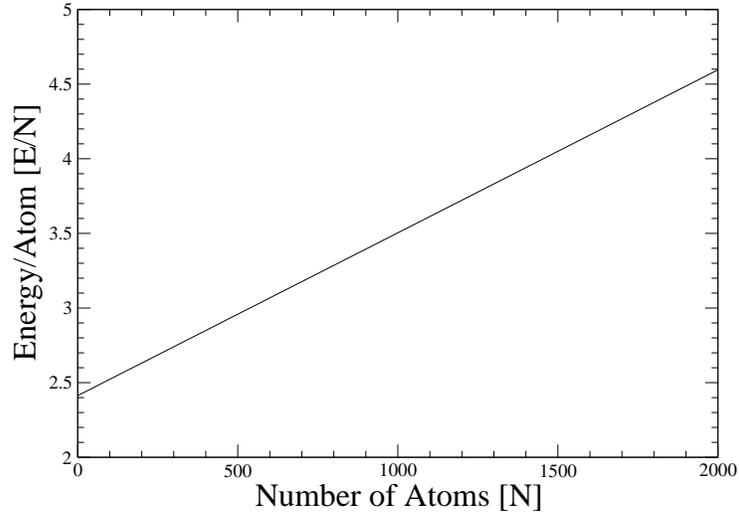}}
\caption{A plot for the Condensate Energy/Atom  versus Number of atoms in 
trap for 2000 particles for the ground state  
; this work}
\end{figure}
\begin{figure}[h!]
\centering
\epsfxsize=3.2in{\epsfbox{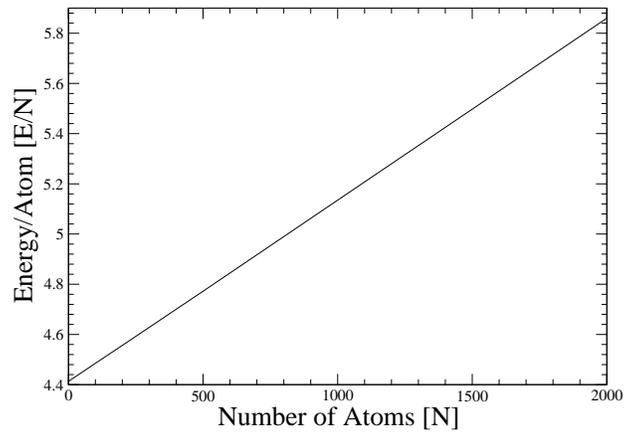}}
\caption{A plot for the Condensate Energy/Particle  versus Number of atoms in 
trap for 2000 particles for the 
 $m=2$ mode; this work}
\end{figure}
\vskip 2cm
\newpage
\begin{table}[h!]
\caption{\bf frequency $\omega$ for lowest lying modes}
\begin{center}
\begin{tabular}{|c|c|c|c|c}
\hline
N & mode of oscillation & energy & $\omega$(this work) & $\omega$(JILA TOP) \\
\hline\hline
2000 & ground state & 4.596(3)\\
2000 & $ m=2$& 5.860(1) &1.264(4)& 1.4 \\ 
2000 & $m=0$ &7.295(3)& 2.699(6)&1.8 \\
\hline
\end{tabular}
\end{center}
\end{table}
\vskip 2cm
\begin{figure}[h!]
\centering
\epsfxsize=4in{\epsfbox{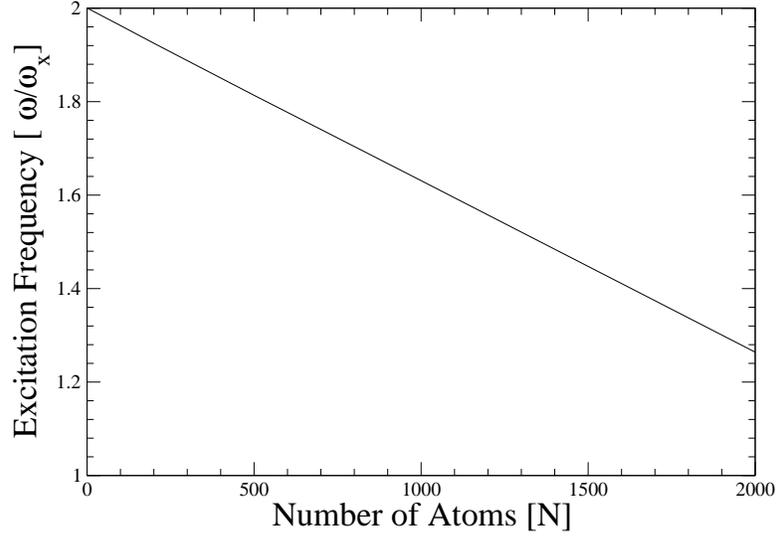}}
\caption{ A plot of Excitation Frequency vs Number of Atoms 
for lowest lying $ m=2 $ mode ;this work}
\end{figure}
\begin{figure}[h!]
\centering
\epsfxsize=4in{\epsfbox{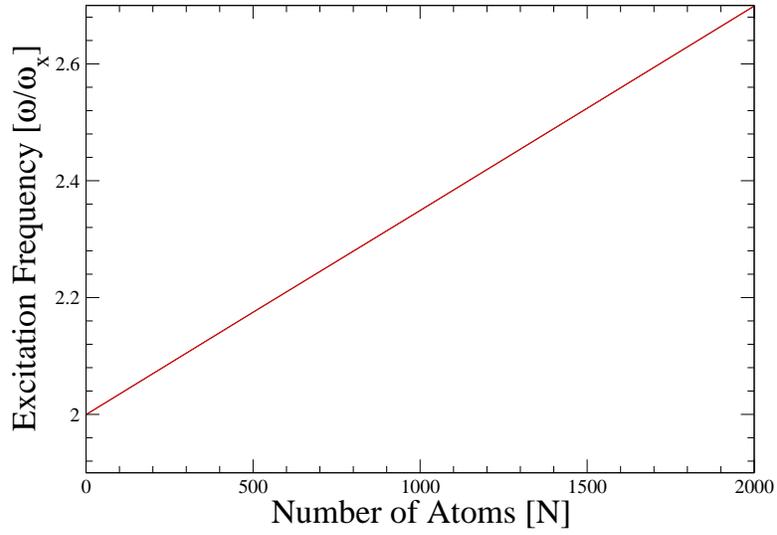}}
\caption{ A plot of Excitation Frequency vs Number of Atoms
for lowest lying $ m=0$ mode ; this work}
\end{figure}

\begin{figure}[h!]
\centering
\epsfxsize=4in{\epsfbox{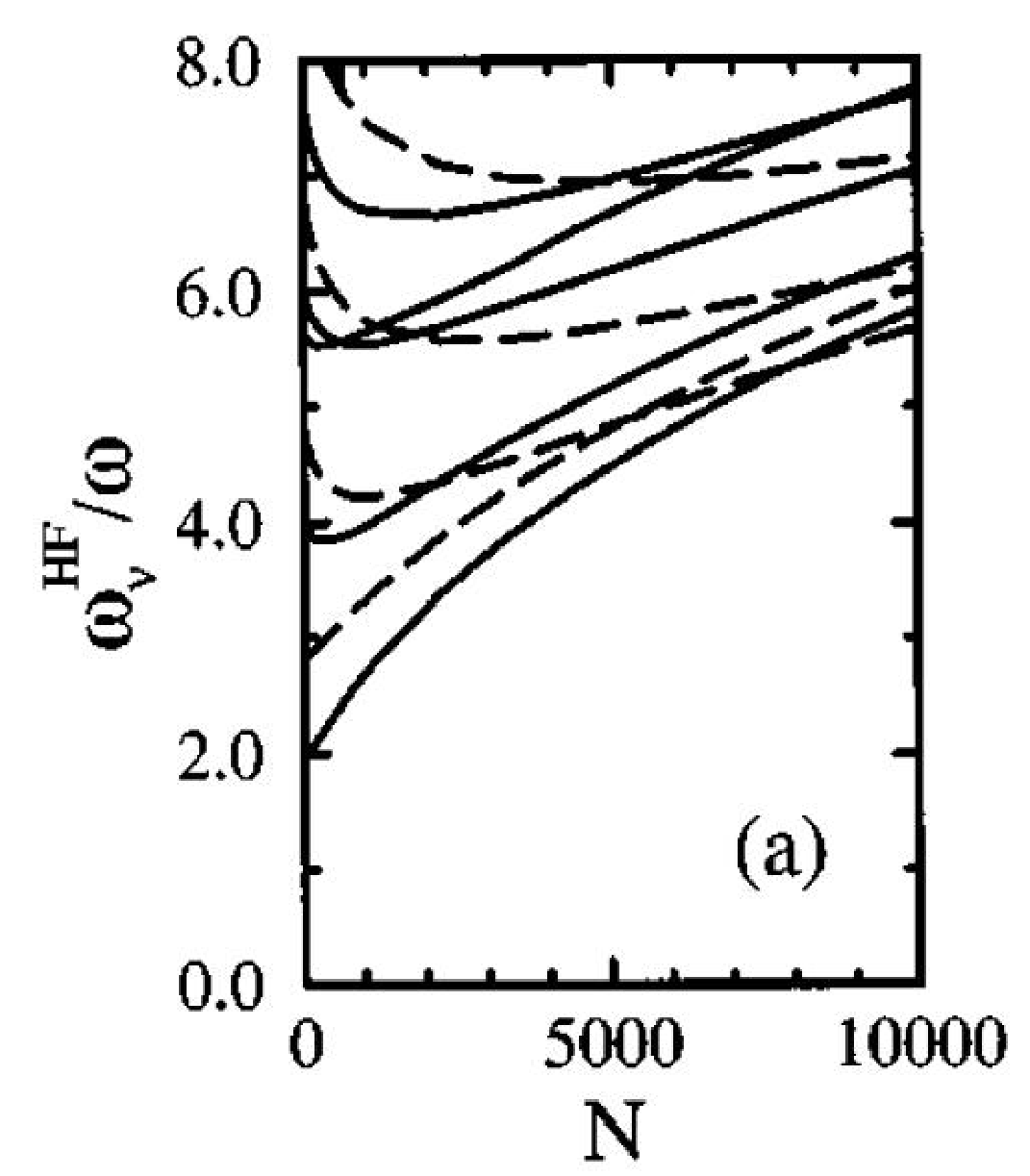}}
\caption{frequencies for m0;Ref 33}
\end{figure}
\newpage
\subsection{Effects of temperature on condensation fraction}
Density of condensate atoms decreases in the trap as temperature increases.
This lowers the interaction energy of the condensate atoms resulting in a shift
in the critical temperature.  As a matter of fact in the interacting case, 
the critical temperature decreases. This is a very unique feature of trapped 
gas. In the case of uniform gas we see an opposite trend. 
\begin{figure}[h!]
\centering
\epsfxsize=3.5in{\epsfbox{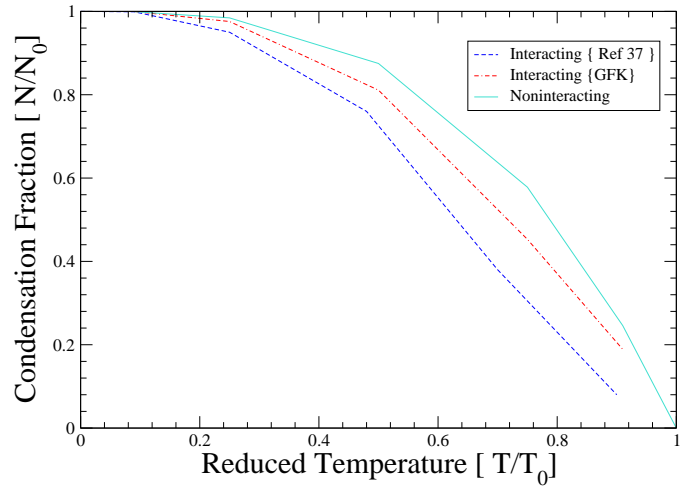}}
\caption{Condensation fraction vs Reduced Temperature
; this work. The middle curve corresponds to the 2000 interacting atoms(GFK sumlation data;ourwork) and 
the outer one corresponds to the noninteracting case. The innermost curve also
corresponds interacting atoms[Ref 37].The number of condensed 
particles decreases with the interaction}.
\end{figure}
\newpage
\subsection{Effects of temperature on the frequency shifts; 
comparison with other experiments and theories}
Next we give an account of how our path integral simulations compare with 
other theoretical and experimental data. Fig 3a in Ref[2] represents JILA TOP 
data where  one observes a large temperature dependent 
frequency shift for both m=0 and m=2 modes. For m=2 mode, starting from 
Stringari limit it decreases all the way up to $0.9T_0$ whereas for m=0 mode 
it shows a rising trend with rise in temperature. Our path integral data for 
$m=2$ mode in Fig[6] shows similar decreasing trend as  JILA TOP data 
in Fig 3a of Ref[2] and best theoretical 
data in Fig 1 of Ref[7] all the way up to $0.9T_0$ whereas data generated 
by Hartree-Fock-Bogoliubov[HFB] method in Fig 1 of Ref[4] agree with 
JILA TOP only up to $0.7T_0$ .
Finally our data Fig[7] for  
temperature variation of  m=0 mode 
 agrees with JILA data in Fig[3a] of Ref[2] and Morgan data in Fig 1 of Ref[7] 
when the effect of thermal cloud  is considered.
\begin{figure}[h!]
\centering
\epsfxsize=3.0in{\epsfbox{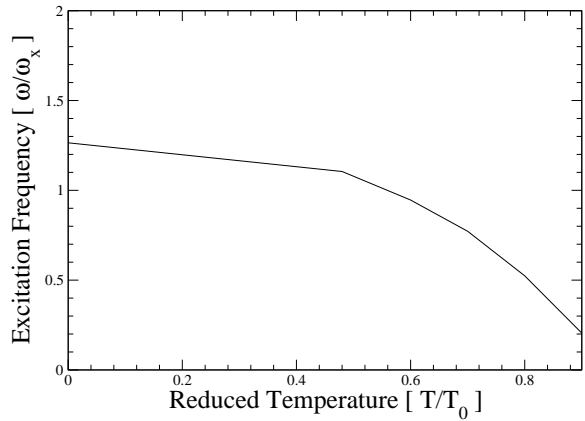}}
\caption{Effects of temperature on $m=2$ mode; this work}
\end{figure}
\begin{figure}[h!]
\vskip 1cm
\centering
\epsfxsize=3.1in{\epsfbox{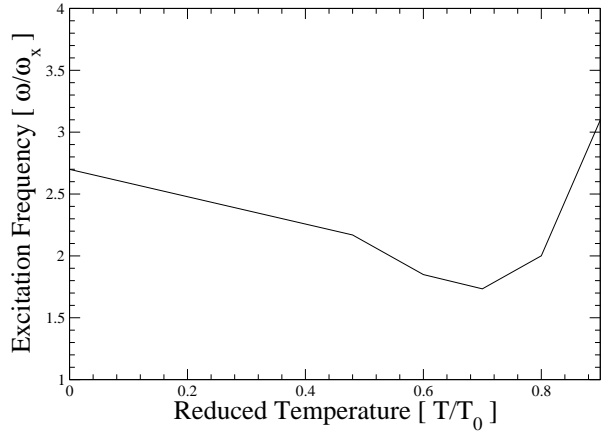}}
\caption{Effects of temperature on $m=0$ mode from GFK considering
noncondensate dynamics[this work], shows resemblence with JILA[2] and 
Morgan et al[7]}
\end{figure}
\newpage
\subsection{Discussions}
 We are solving the full Hamiltonian with some realistic potential and thereby
trying to solve the many body problem fully quantum mechanically and
nonperturbatively. So for both $m=0$ and $m=2$ modes, our calculations 
adopting Feynman Kac path integral technique represent the collective behavior
of the Bose gas. Our work [Fig. 6] agrees with JILA experiment[2] for m=2 mode. 
The other theoretical work shows the reverse trend [Fig 1 of Ref 4]. 
We found that considering the dynamics of condensates alone and
the effect of finite temperature as static thermal cloud, we do not achieve the
upward shifts of frequencies as shown by JILA data for m=0 mode.
In fact, we agree with Ref [5] that as m=0 happens to be a coupled
mode, we need to consider the dynamics of thermal cloud to obtain a
satisfactory agreement with the experimental data. Eventually We  consider
the dynamics of thermal cloud and get the upward shift of data[Fig 7] as
observed in JILA[2] and Ref[7,8].

At T=0 for $m=2$, we observe that as N increases
the energies grow, but the splitting between the ground and excited state
decreases - an essential feature of Bose Condensation. But  in the similar 
case for $m=0$ mode we observe that both the energies and the gap between the 
ground and higher excited states increase with increase in number of atoms.This
 agrees with Hartree-Fock spectrum of Ref[33] where the author had to improve
these results by using random phase approximation to get an agreement with the 
experimental results. For $m=2 $ mode the Stringari limit turns
out to be 1.264(4) as opposed to the experimental value of  1.4. On the other 
hand $m=0$ mode the Stringari limit is 2.699(6) as opposed to the experimental
 value of 1.8.
So for the coupled $m=0$ mode, our path integral method
 does not work better than Hartree-Fock theory and thereby does not yield
a correct value of Stringari limit. The reason that the quantitative 
agreement between  the Stringari limit predicted by us particularly for $ m=0$ 
mode  and the experiments is not good, 
might be the use of Gaussian wave functions as the trial functions. In
general, the frequencies of collective modes do not have direct correspondence
to harmonic oscillator states because they do not include correlations.
It is legitimate to use harmonic oscillator solutions as trial functions
 for m=2 mode as JILA experiment does not correspond 
to the Thomas Fermi limit[35]. But for m=0 mode we need to include 
correlations in the wave function as m=0 is a coupled mode.
\newpage 
\section{Conclusions:}
We have used GFK to bring out the many body effects between the cold Rb atoms.
Numerical work with bare Feynman-Kac procedure employing modern computers was
reported[15] for the first time for few electron systems after forty years of
the original work[14] and seemed to be really useful for calculating atomic 
ground states[19]. A fairly good success in atomic physics motivated us to 
apply it to Condensed matter Physics.

Gross-Pitaevskii (GP) technique[33] does not include correlations in the 
solutions explicitly and calculates energy at the variational level.
These energies are upper bounds to the actual energies of the system.
We have been successful in achieving a lower value for Rb ground state than 
that obtained by GP.  This correct trend in 
our calculated energies for different symmetry states enables one to calculate 
the frequencies more accurately by Feynman Kac path integral method. 
 As a result, Diffusion Monte Carlo codes
based on nonperturbative quantum appraoach can handle temperature very 
accurately and we do not see any breakdown near $T_c$. For the first time we 
have calculated finite temperature properties beyond mean field approximation 
by Quantum Monte Carlo technique. The only other non mean field calculations 
at $T=0$ worth mentioning in this context is the work done by Blume et al[36].
 We have calculated spectrum of Rb gas by 
considering realistic potentials like Morse potential etc. instead of 
conventional pseudopotentials for the first time. 

We have been able to calculate the lowest lying excitation frequencies for
$m=0$ and $m=2$ modes by Feynman-Kac path integral technique in a very simple
way. We have found an alternative to Gross-Pitaevskii technique and other mean 
field calculations which works at all the temperature.
Simulating 2000 atoms with the path intgral method we have been able to 
capture some of the signatures of Bose condensation like decrease of excitation
frequencies with number of atoms, lowering of condensation fraction in the interacting case etc.
In our non mean field study, we see 
agreement with experimental study all the way to $ T=0.9T_c $ [Fig. 6]. 
This is because of the fact that we have been able to 
solve the related many body theory very accurately with the nonperturbative 
and quantum mechanical approach. At this point our results agree with the \
experimental results only qualitatively as we are restricting ourselves 
to the choice of Gaussin trial functions. To improve our results 
quantitatively we need to use correlated trial functions. This would be a
nontrivial extension of the present work and will be reported elsewhere.
The simplicity in our method is appealing as it is  extremely easy to 
implement and our fortran code at this point consists of about 270 lines. 
In fact mere ability to add, subtract and toss a coin enables one to solve 
many body theory with our path integral technique. 

We employ an algorithm which is essentially parallel in nature so that 
eventually we can parallelize our code and calculate thermodynamic properties 
of bigger systems  taking advantage of new 
computer architechtures. This work is in progress. We are continuing on this 
problem and hope that this technique will inspire others to do similar 
calculations. 

\newpage
{\bf Acknowledgements}:\\
Financial help from the Department of Science and Tecnology(DST), India 
( under Women Scientist Scheme; award no. SR/WOS A/PS-32/2009 ) is gratefully acknowledged. 
\end{document}